\documentclass[12pt]{iopart}
\usepackage{graphicx,subfigure}
\usepackage{bm}        
\usepackage[mathscr]{euscript}
\begin{document}
\newcommand{\braket}[3]{\bra{#1}\;#2\;\ket{#3}}
\newcommand{\projop}[2]{ \ket{#1}\bra{#2}}
\newcommand{\ket}[1]{ |\;#1\;\rangle}
\newcommand{\bra}[1]{ \langle\;#1\;|}
\newcommand{\iprod}[2]{\bra{#1}\ket{#2}}
\newcommand{\logt}[1]{\log_2\left(#1\right)}
\def\cI{\mathcal{I}}
\def\cC{\tilde{C}}
\newcommand{\cx}[1]{\tilde{#1}}

\def\be{\begin{equation}}
\def\ee{\end{equation}}
\title[Loschmidt echo of local dynamical processes in spin chains]{{Loschmidt echo of local dynamical processes in integrable and non integrable spin chains}}

\author{Saikat Sur$^1$ and V. Subrahmanyam$^2$}
 \address{ Department of Physics, Indian Institute Of Technology,  Kanpur-208016, India}
 
 \ead{saikatsu@iitk.ac.in$^1$ and vmani@iitk.ac.in$^2$}
\date{\today}
\begin{abstract}
The Loschmidt echo is investigated to track the effect of the local QDP. It is also quite sensitive to
whether the background dynamics is integrable or not. For the integrable case, viz. the Heisenberg
model, the Loschmidt echo depends on the parameters operators corresponding to the QDP as well as the time of QDP. The probability of reviving the system to its initial state is higher for incoherent QDPs occurring at large time intervals. Whereas each time coherent QDP occurs certain probability of reviving the state is always lost. For  For the non-integrable case, viz. a kicked Harper model,
it exhibits a decaying behaviour when contrasted with integrable dynamics. The decay rate is slower when the corresponding classical Hamiltonian is non chaotic. The Loschmidt echo also distinguishes the integrable and the nonintegrable dynamics when a QDP occurs. 
 
\vskip 0.5cm
\hfill {PACS: 03.65.Ud 03.67.Bg 03.67.Hk 75.10.Pq}
\end{abstract}
\maketitle

 \maketitle
\section{ Introduction}

An isolated system evolves unitarily which allows a pure state to remain pure through out the evolution. So, such systems can be brought back to their initial state through time reversal operation. If the system interacts with the environment this is no longer true and such systems cannot be brought back to their initial state by time reversal operation. Moreover, it has been proposed that the probability of restoring the system back to the initial state is a decreasing function of time. In addition,  the revival probability shows a contrasting behaviour for integrable and non integrable background dynamics. Peres \cite{peres} first showed that classically chaotic and integrable system behave differently under imperfect time reversal. The Loschmidt Echo is a measure of the revival of the state when an imperfect time reversal procedure is applied during the evolution of a quantum system. It quantifies the sensitivity of the quantum state to the local decohering process when the system interacts with the environment.\\

 The Loschmidt Echo in many body system has been studied extensively over the last decade\cite{gorin}. It has been studied in contexts like quantum quench \cite{happola,jafari, torres}, sensitivity to a perturbation in many body systems \cite{elsayed}, many body localised phase \cite{serbyn}, in connection with the Lyapunov exponents of classical systems \cite{cpj},many body system in a decohering environment \cite{quan,sharma, zangara2, zangara3}.  It has been observed that the Loschmidt Echo decays exponentially in ergodic systems \cite{serbyn,torres,iomin}.
 The time scale of the decay can be related to the Lyapunov exponent of the classical system \cite{fine,serbyn,cpj}.\\ 

A system that is undergoing a unitary evolution due to a Hamiltonian dynamics can be interrupted by a quantum processes,  operating locally and instantaneously, and further evolution through the background Hamiltonian dynamics.
The local operation that interrupts the background evolution can occur from local quantum decohering processes or local coherent operations. 
The details of the effects of 
A local quantum dynamical process (QDP)  occurring on a many qubit state during unitary evolution, can change the distribution of
correlations and entanglement structure. The speed of the signal propagating due to the QDP occurrence has been studied\cite{ssvs}, and the interference of the signal and the state propagation in the spin chain dynamics has been investigated recently\cite{ssvs2}. The effect of a local quantum dynamical process (QDP) at a given qubit at a given epoch of time during the unitary evolution of an initial state can also be studied using the Loschmidt echo, calculated from the overlap of the time evolved multi-qubit states with and without QDP.\\

In this paper we will concentrate on both coherent and incoherent QDPs. An incoherent QDP will cause decoherence in the system, and the state will become a mixed state as a result. Multiple incoherent QDPs intervening the dynamics at regular intervals can be thought of as if the system is interacting with an external decohering environment. Multiple number of coherent QDPs although do not cause decoherence but can generate non integrability in the system. Certain non integrable systems described by time-dependent Hamiltonians can be thought of as a multiple coherent QDPs intervening an integrable dynamics at different qubits with different strengths. We will discuss how non-integrable dynamics can be obtained by introducing multiple coherent operations at a regular interval on integrable dynamics. We will see below that the  Loschmidt Echo is able to show the contrasting behaviour between the integrable and non-integrable dynamics.

The paper is arranged as follows. In Section 2,  we will discuss a general approach to compute the Loschmidt Echo for anisotropic Heisenberg dynamics for both incoherent and coherent QDPs using Green's functions. We will also consider the case of multiple number of QDPs occurring on the spin chain. In Section 3,  we will study the Kicked Harper model dynamics, that allows us to go from an integrable to a non-integrable dynamics by turning the kicking time. We will investigate the contrast between the integrable XY dynamics and Harper dynamics, chaotic and non chaotic Harper dynamics, and incoherent QDP intervening the Harper dynamics by computing the Loschmidt Echo.  In the last section, we give a summary and conclusions.

\section{Local QDP interrupting the dynamics of Heisenberg model}

   We consider here the time evolution of a Heisenberg spin chain starting with an initial state $|\Psi(0)\rangle$ at time $t=0$. The background Hamiltonian dynamics is interrupted due to QDP occurring at time $t=t_0$ during the evolution of the state from time $t=0$ to time $t$.  Thus, the time evolution  here occurs in three stages: The first stage is a unitary evolution from $t=0$ to the epoch of QDP at time $t=t_0$, represented by a unitary operator $U_{t_0,0}$. The second stage is the action of an instantaneous QDP at $t=t_0$, which will be represented by Kraus operators for an incoherent QDP, and a unitary operation for coherent QDP.  The third stage is again a unitary evolution from time $t=t_0$ to time $t$, represented by the unitary operator $U_{t,t_0}$.
In this case after the local instantaneous local operation on the system will not be able to revive to the initial state if the time reversal process is performed using the background dynamics,
which means the Loschmidt echo may not be equal to unity. In our context the Loschmidt Echo is defined as the probability of the state to return to its initial state by time reversal unitary process after the operation which occurs at $t_0$. It is quantified by taking overlap of the time reversal state with the initial state as a function of time. In our case, the state evolves unitarily after the  incoherent QDP the  Loschmidt Echo will not be a function of time $t$  but a function of $t_0$, the time of QDP occurrence and in general the location index $m$ where it occurs.  In this section, and in the next section, we will study the dynamics of many qubit systems (both integrable and non-integrable) under local instantaneous operations   

The Loschmidt echo, for unitarily evolved states, is  just the overlap of the two states evolved under two different Hamiltonians. In our case, the evolution with a QDP can lead to a
mixed state, represented by $\tilde \rho(t)$. The state evolved without a QDP remains a pure state,  represented by $\rho(t)=|\Psi(t)\rangle \langle \Psi(t)|$.
Thus, the Loschmidt Echo  $L(m,t_0)$ can be defined as,  
   \begin{equation} 
{L}(m,t_0) =Tr \tilde \rho(t) \rho(t) =\langle \Psi(t) | \tilde \rho(t)|\Psi(t)\rangle,
\end{equation}
where  $m$ keeps track of the location of QDP, and $t_0$ the time of QDP occurrence.
Though, the overlap of the two states is taken at time $t$, the Loschmidt echo will depend only on $t_0$, as the two evolutions are same except at $t=t_0$. A decohering or incoherent QDP
is a non-unitary operation,  for example the action of a single-qubit quantum channel. The Loschmidt Echo can in general depend on the initial state,  the Kraus operators  of the QDP,  the time of the QDP, and the site of the QDP.  

The state of the system,  for a time $t<t_0$ before the QDP occurs is given by the unitarily evolved state given by $\rho(t)=|\Psi)t)\rangle\langle \Psi (t)|$, from the initial state $\rho(0)$, where the state $|\Psi (t)\rangle= U_{t_0,0} |\Psi(0)\rangle$. Just after the instantaneous action of the QDP, we can represent the state using the Kraus operators  $\{E_i\}$, that represent the dynamical process. 
After the QDP occurs at $t=t_0$ at a site $m$ the state can be written as,
\begin{equation}
\tilde \rho(t_0^+) = \sum_{i} E_i \rho(t_0) E^{\dagger}_i .
 \end{equation}  
 Further evolution of the state is a unitary evolution from $t_0$ to an arbitrary time $t$, the state is given by $\tilde \rho(t)=U_{t,t_0}\tilde \rho(t_0) U_{t,t_0}^\dag$.
 Now, the expression for the Loschmidt Echo can be rewritten in terms of the Kraus operators, is  given by,
 \begin{eqnarray}
{L}(m,t_0) = \sum_i |\langle \Psi(0)| U^{\dagger}_{t_0,0} E_i  U_{t_0,0}|\Psi(0)\rangle|^2  = \sum_i |\langle \Psi(t_0)| E_i |\Psi(t_0)\rangle|^2.
\end{eqnarray}  
Some  examples of single qubit operations are  a phase flip gate, a  bit flip gate , a projective measurement in the eigen basis of $\sigma^z$. 
The Kraus operators for a QDP which is a phase flip gate are given by $E_0 = \sqrt{p} \mathbf{1}, E_1 = \sqrt{1-p} \sigma^z$. Similarly for a bit flip gate, the Kraus operators are given by
S $E_0 = \sqrt{p} \mathbf{1}, E_1 = \sqrt{1-p} \sigma^x$. For a measurement process along z-axis, the Kraus operators are given by  $E_0 =(1+\sigma^z)/2, E_1 = (1-\sigma^z)/2$. For all these processes, the Loschmidt Echo has similar form as given above. 
For a phase-flip gate QDP acting on the spin at site $m$ at time $t_0$,  the expression for $L(m,t_0)$  is given by,
 \begin{eqnarray}
{L}(m,t_0) = p + (1-p)|\langle \Psi(t_0)| \sigma^z_m |\Psi(t_0)\rangle|^2.
\end{eqnarray}

To compute the Loschmidt echo, we need the time-evolved state under the background Hamiltonian dynamics, and the expectation value of the Kraus operators in the state.
 We will first discuss the effect of local operations (both coherent and incoherent) in Heisenberg model, an exactly solvable and integrable model. Let us consider a one dimensional chain of $N$ spin interacting with its nearest neighbour. The Hamiltonian is given by,
\begin{equation}
H = -{1\over 2} \sum_i \big(\sigma^x_i \sigma^x_{i+1} + \sigma^y_i \sigma^y_{i+1} +\Delta \sigma^z_i \sigma^z_{i+1}\big).
\end{equation} 
Using the $\sigma^z$ basis states (up-spin and down-spin states for each spin), the last term signifies the interaction between spins, and the first two terms signify the hopping of down spins or up spins. 
The total number of up(down) spins or z component of the total spin is a constant of motion in the dynamics. Thus, the eigenstates have definite number 
of up(down) spins. The ground state is a ferromagnetic state with all the spins up or down, with the energy $\epsilon_g= -N \Delta/2$ for a chain of $N$ 
spins. One-magnon excitations can be created by turning any one of the spins, giving $N$ localised one-magnon states, which can be labelled by the location of the down spin. The state for one down spin at the site $x$ can be represented by  $|x\rangle = \sigma^-_x|F\rangle$. $|F\rangle$ is the ferromagnetic ground state of the Hamiltonian. One-magnon eigenstates are labelled by the momentum of the down spin,  with a plane-wave eigenfunctions. The one-magnon eigenvalue is given by $\epsilon_1(p)=\epsilon_g-2\cos{p}$, where the momentum $p=2\pi l/N$ is determined by an integer $I=1,2,..N$. The interaction of the two down spins is determined by $\Delta$. 

For our discussion below, we will consider two types of initial states.  Consider an unentangled initial state  $\alpha |F\rangle + \beta |1\rangle$, which is a linear combination of the ferromagnetic ground state and a 
one-magnon state with the down spin localised on site 1, This is an unentangled state, being a direct product of up-spin states for all sites and a linear combination of up and down-spin states for the first spin.  For later times, the one-magnon component evolves, through spreading of the down spin wave function. We will also consider an entangled state,  a one-magnon initial state $\alpha |1\rangle + \beta |r\rangle$, where the first and the $r^{th}$ qubits are entangled initially, the rest of the spins are in a direct product state. For later times, the down-spin wave function will evolve by spreading out, generating entanglement for other pairs of spins.

For the unentangled initial state $|\Psi(0)\rangle=\alpha |F\rangle + \beta |1\rangle$,  the time-evolved state can be written using the one-magnon propagator  as
\begin{equation}
|\Psi(t)\rangle=e^{-i\epsilon_gt} \big [ \alpha |F\rangle +\beta \sum_x G_1^x(t)|x\rangle \big ],
\end{equation}
where $E_g$ is the ground state energy, and the one-magnon Green's function is given in terms of the eigenfunctions and eigenvalues given above, defined by  
   \begin{equation} 
 G^{x'}_{x}(t)=e^{i\epsilon_gt}\sum_{p}\psi_{p}^x  \psi_p^{x'*} e^{-it\epsilon_1 (p)}.
  \end{equation}
  Using the one-magnon eigenfunctions we can express the Green's function\cite{ssvs} for a large systgem as,
   \begin{equation} 
 G^{x'}_x(t)  = (-i)^{(x-x')}J_{x-x'}(2t),
  \end{equation}  
  where $J_n(2t)$ is the $n$'th order Bessel function.
We can evaluate the expectation values of the spin operators at later times using the one-magnon Green's function,
we have
    \begin{eqnarray}
  &\langle \Psi(t_0)|\sigma_m^z|\Psi(t_0)\rangle = 1-2|\beta|^2 |G^m_1(t_0)|^2, \nonumber\\
  &\langle \Psi(t_0)|\sigma_m^x|\Psi(t_0)\rangle = 2 \alpha^* \beta Re( e^{-i\epsilon_0 t_0}G^{m}_1(t_0)).\nonumber\\
 \end{eqnarray}
 Hence, we get an expression for the Loschmidt Echo for the phase-flip QDP as, 
  \begin{eqnarray}
  &L(m,t_0) = p+(1-p)(1-2|\beta|^2 |G^m_1(t_0)|^2)^2.
   \end{eqnarray}
   Similarly, we can evaluate the Loschmidt echo for for a bit-flip gate QDP, we have
   \begin{eqnarray}
  &L(m,t_0) = p+4|\alpha|^2|\beta|^2(1-p)(Re( e^{-i\epsilon_0 t_0}G^{m}_1(t_0)))^2.
 \end{eqnarray}

   Now, for the entangled initial state $|\Psi(0)\rangle=\alpha |1\rangle + \beta | r\rangle$,  the expectation values  for the spin operators for later times, analogous to Eq.9, are given by,
  
   \begin{eqnarray}
  &\langle \Psi(t_0)|\sigma_m^z|\Psi(t_0)\rangle = 1-2|K^m(t_0)|^2, \nonumber\\
  &\langle \Psi(t_0)|\sigma_m^x|\Psi(t_0)\rangle = 0. \nonumber\\
    \end{eqnarray}
    
  Here,  we have defined a linear combination of the propagators as, $K^m(t) = \alpha G^m_1(t) + \beta G^m_r(t)$.
 Thus,  the expression for the Loschmidt Echo for the phase flip gate is given as, 
 \begin{eqnarray}
  &L(m,t_0) = p+(1-p)(1-2|K^m(t_0)|^2)^2,
     \end{eqnarray}
  and  for the bit flip gate is given by,
   \begin{eqnarray}
  &L(m,t_0) = p.
 \end{eqnarray}

The Green function  appearing in the above equations, since it is a Bessel function with time $t_0$ as argument,  falls inversely with $t_0$. It is easy to see that
 for large values of $t_0$ the $\sigma^z$ expectation value for both entangled and unentangled initial states approaches unity.
 This implies that for a phase-flip QDP or a projective measurement  QDP in $\sigma^z$ basis,  both the states are fully reversible for large value of $t_0$. Now, the $\sigma^x$ expectation value is zero for the entangled state $\alpha |1\rangle + \beta |r\rangle$ for any value of $t_0$, as the operation changes the parity of the state,  and for the unentangled state $\alpha |F\rangle + \beta |1\rangle$, the expectation value  goes to zero for large value of $t_0$. This would imply that,  for the phase-flip gate QDP, and a projective measurement QDP in $\sigma^x$ basis, the Loschmidt Echo tends to $p$ and $1/2$ respectively for the two states for large value of $t_0$; the states are not time reversible  here. These results are shown in Fig 1(a).

 Now, we will consider the case where the quantum system undergoes multiple sequential QDPs during evolution. The QDPs occur at site $m_1$ at time $t_0$, at site $m_2$  at  time $2t_0$ and so on. After two such QDPs at times $t_0$ and $2t_0$ the expression for $L(m_1,m_2;2t_0)$ can be given by extending Eq 4 as,

   \begin{eqnarray*}
  &L(m_1,m_2;2t_0) = p^2 + p(1-p)|\langle \Psi(2t_0)|U_{2t_0,t_0}\sigma^z_{m_1}U_{t_0,0}|\Psi(0)\rangle|^2  \nonumber\\ &+ p(1-p)|\langle \Psi(2t_0)|\sigma^z_{m_2} U_{2t_0,t_0}U_{t_0,0}|\Psi(0)\rangle|^2 + (1-p)^2 |\langle \Psi(2t_0)|\sigma^z_{m_2} U_{2t_0,t_0}\sigma^z_{m_1}U_{t_0,0}|\Psi(0)\rangle|^2. \nonumber\\
    \end{eqnarray*}

 The expression for the Loschmidt Echo after n sequential QDPs at sites $(m_1,m_2,..,m_n)$ at a time interval of $t_0$  is given by,

  \begin{figure*}[t]
     \begin{center}
        \subfigure[]{%
            \label{fig:first}
            \includegraphics[width=0.4\textwidth]{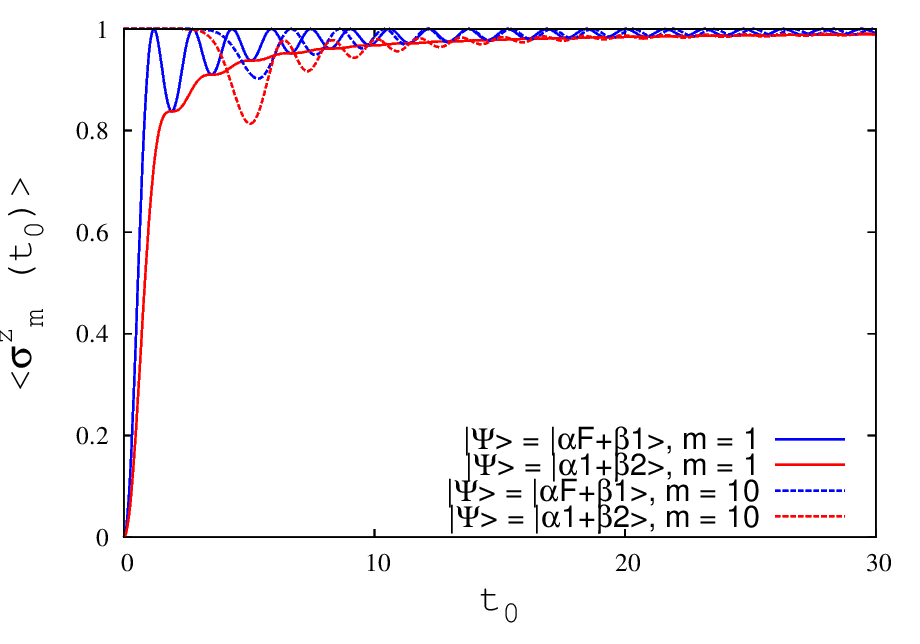}
        }%
        \subfigure[]{%
           \label{fig:second}
           \includegraphics[width=0.4\textwidth]{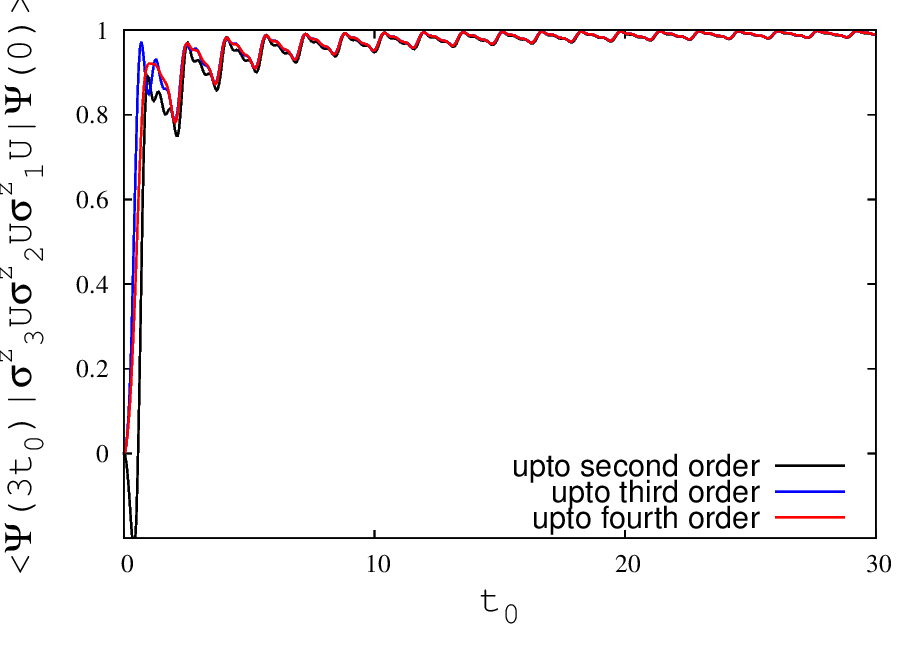}
        }\\
        \subfigure[]{%
            \label{fig:first}
            \includegraphics[width=0.4\textwidth]{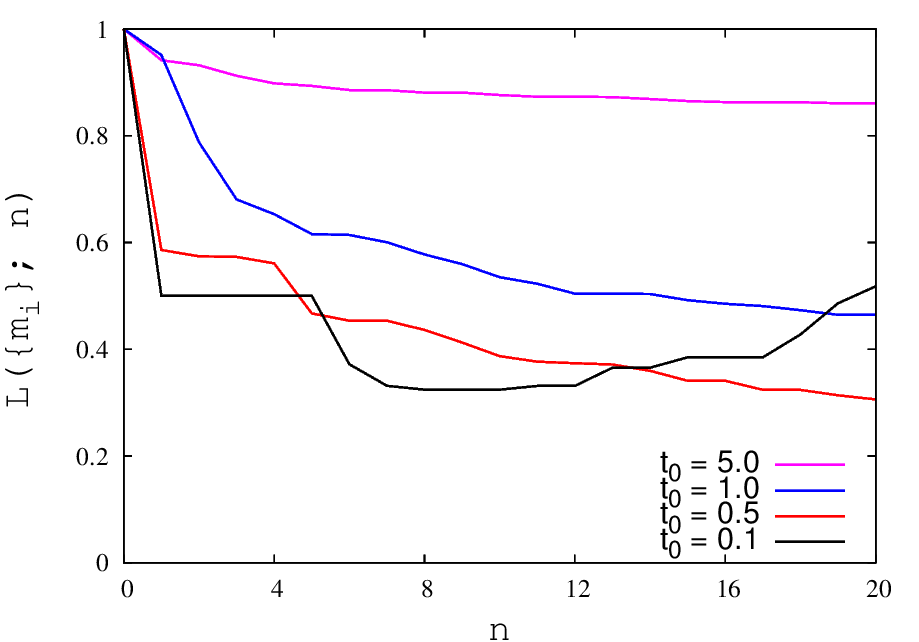}
        }%
        \subfigure[]{%
           \label{fig:second}
           \includegraphics[width=0.4\textwidth]{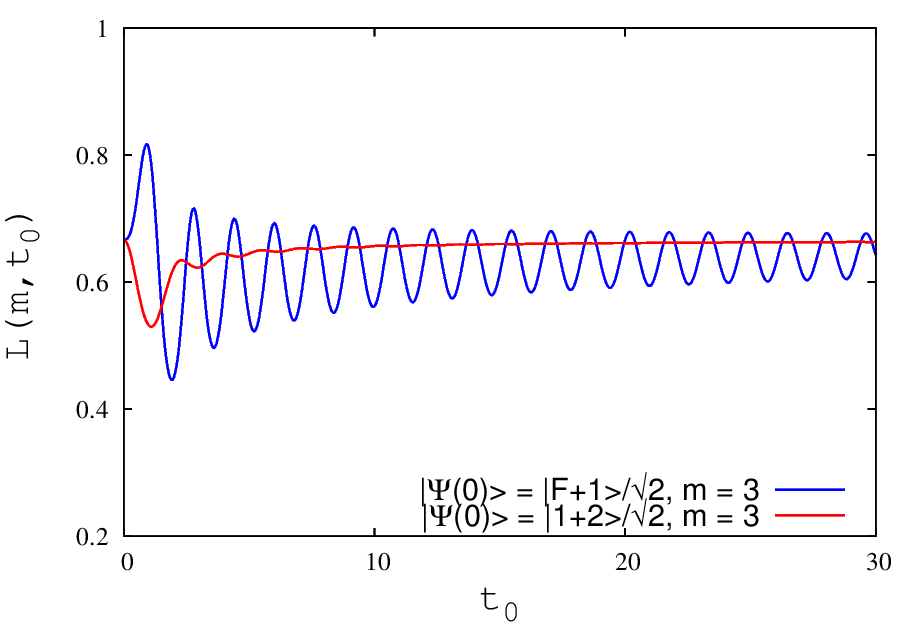}
        }\\
    \end{center}
\caption{\label{fig:fig_1}{(a)The Loschmidt Echo $L(t_0)$ as a function of time of QDP $(t_0)$ for Heisenberg model for Projective measurement along $\sigma^z$ and $\sigma^x$ basis for two different initial states. (b) The quantity $\langle \Psi(3t_0)|\sigma_{3}^zU\sigma_{2}^zU\sigma_{1}^z U|\Psi(0)\rangle$ from Eq 16 for the initial state $\alpha |F\rangle+ \beta|1\rangle$ as a function of $t_0$ upto second order, third order and fourth order terms of the Green function. (c) The Loschmidt Echo for multiple QDPs $L(\{m_i\};n)$ for sequential projective measurement along $\sigma^z$ basis for $t_0 = 0.1,0.5,1.0, 5.0$ as a function of number of QDPs $n$ plotted from Eq 15, taking upto quadratic terms of Green functions. The set of $\{m_i\}$ is taken from random numbers between 1 and 9. (d) The Loschmidt Echo $L(t_0)$ as a function of time of QDP $(t_0)$ for Heisenberg model for coherent QDP for two different initial states, the parameters for coherent QDP are taken as$\gamma = \frac{1}{\sqrt{3}}(1+i)$, $\delta =  \frac{1}{\sqrt{3}}$ from Eq 23 and  Eq.24. }
     }%
   \label{fig:subfigures}
   
\end{figure*}
  
 \begin{eqnarray}
&{L}(\{m_i\};nt_0) = p^n +p^{n-1}(1-p)\sum\limits_j|\langle\Psi(jt_0)| \sigma^z_{m_j} |\Psi(jt_0)\rangle|^2
 \nonumber\\ &
 + p^{n-2}(1-p)^2 \sum\limits_{j,k}|\langle\Psi(kt_0)| \sigma^z_{m_k} U_{kt_0,jt_0}\sigma^z_{m_j} |\Psi(jt_0)\rangle|^2
  \nonumber\\&
  ..+(1-p)^n |\langle\Psi(nt_0)|\sigma^z_{m_n}U_{nt_0,(n-1)t_0}..\sigma^z_{m_1}|\Psi(t_0)\rangle|^2. \nonumber\\
 \end{eqnarray}

  \begin{figure*}[t]
     \begin{center}
        \subfigure[]{%
            \label{fig:first}
            \includegraphics[width=0.32\textwidth]{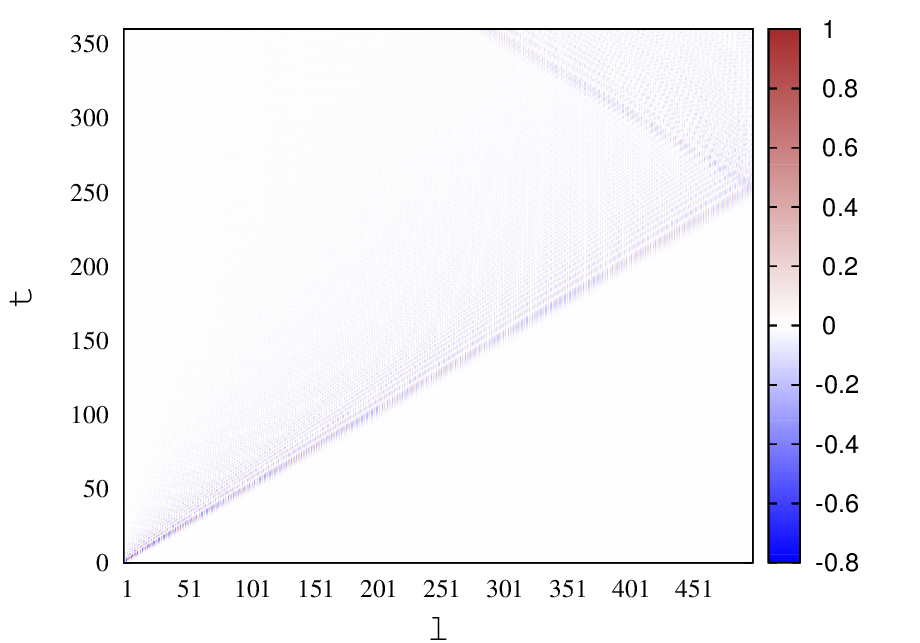}
        }%
        \subfigure[]{%
           \label{fig:second}
           \includegraphics[width=0.32\textwidth]{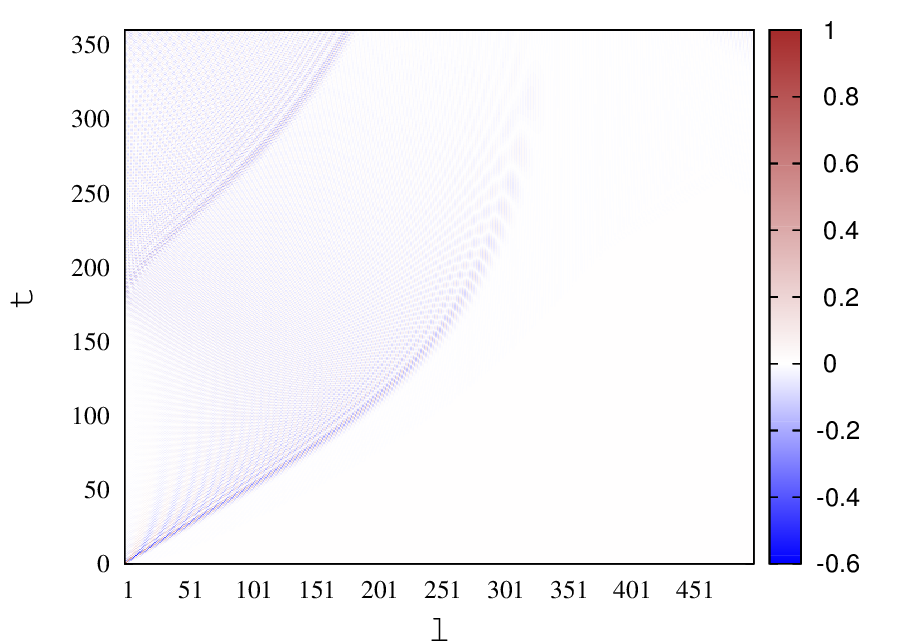}
        }
               \subfigure[]{%
           \label{fig:third}
           \includegraphics[width=0.32\textwidth]{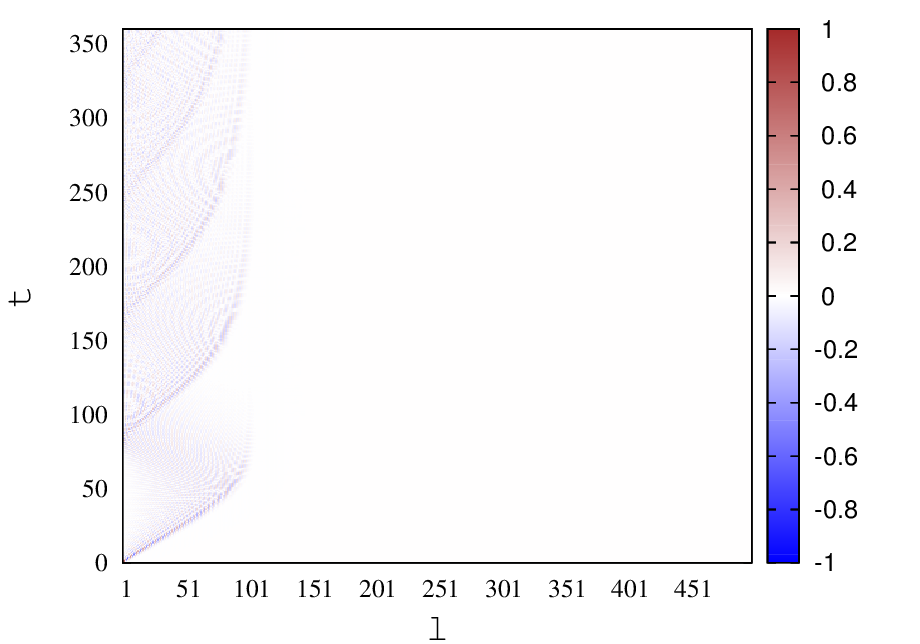}
        }\\%
        \subfigure[]{%
           \label{fig:fourth}
           \includegraphics[width=0.32\textwidth]{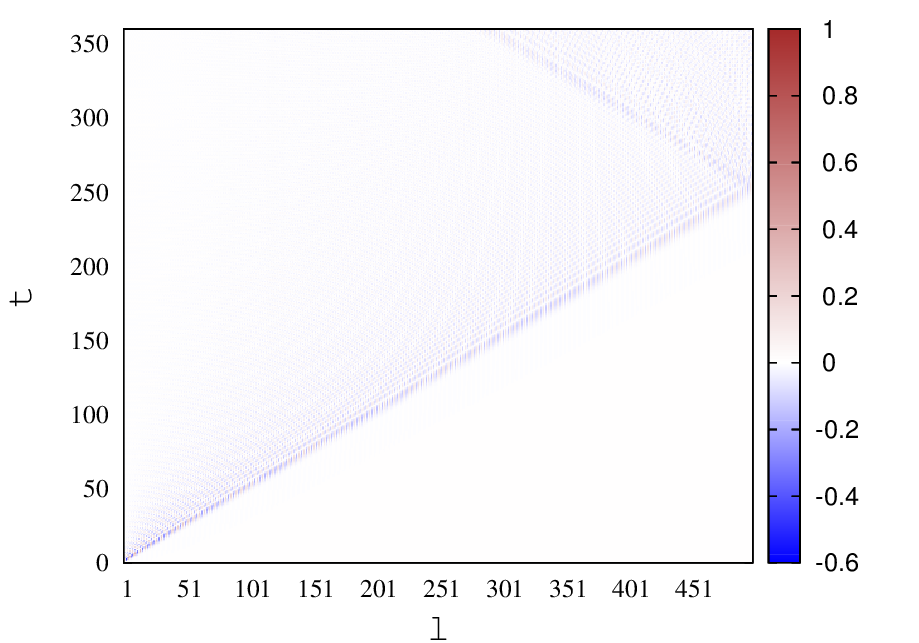}
        }
               \subfigure[]{%
           \label{fig:fifth}
           \includegraphics[width=0.32\textwidth]{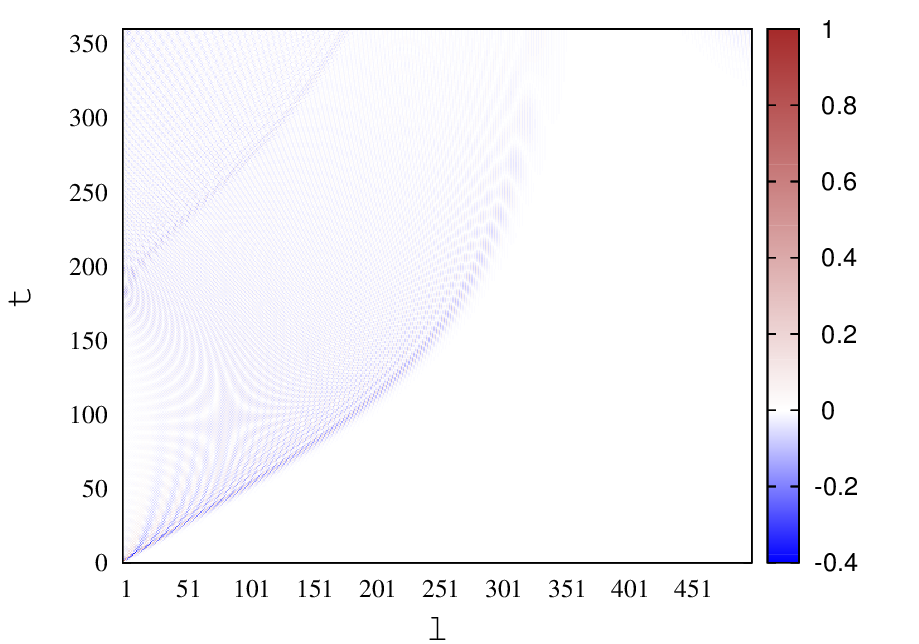}
        }%
        \subfigure[]{%
           \label{fig:sixth}
           \includegraphics[width=0.32\textwidth]{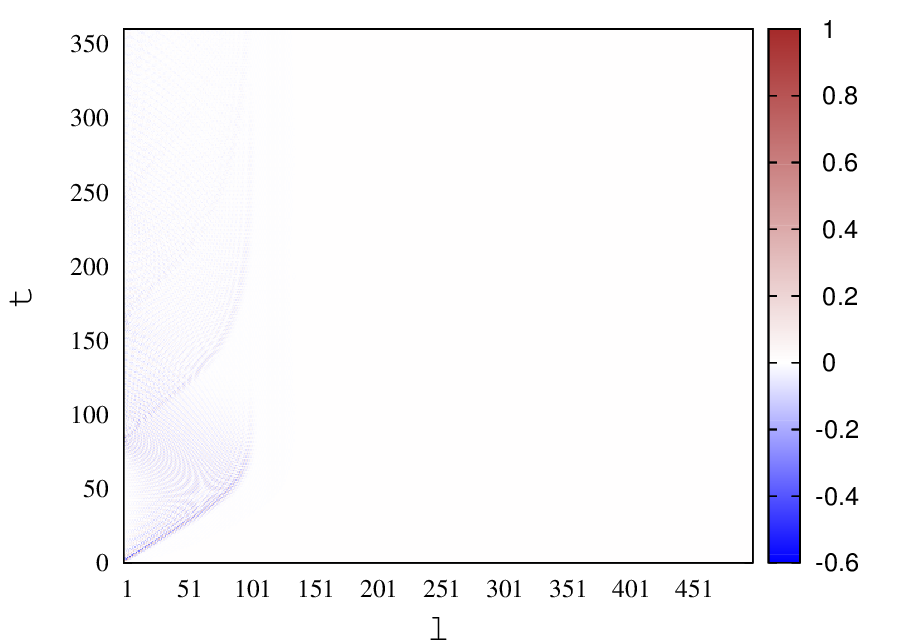}
        }\\
        \subfigure[]{%
           \label{fig:seventhth}
           \includegraphics[width=0.32\textwidth]{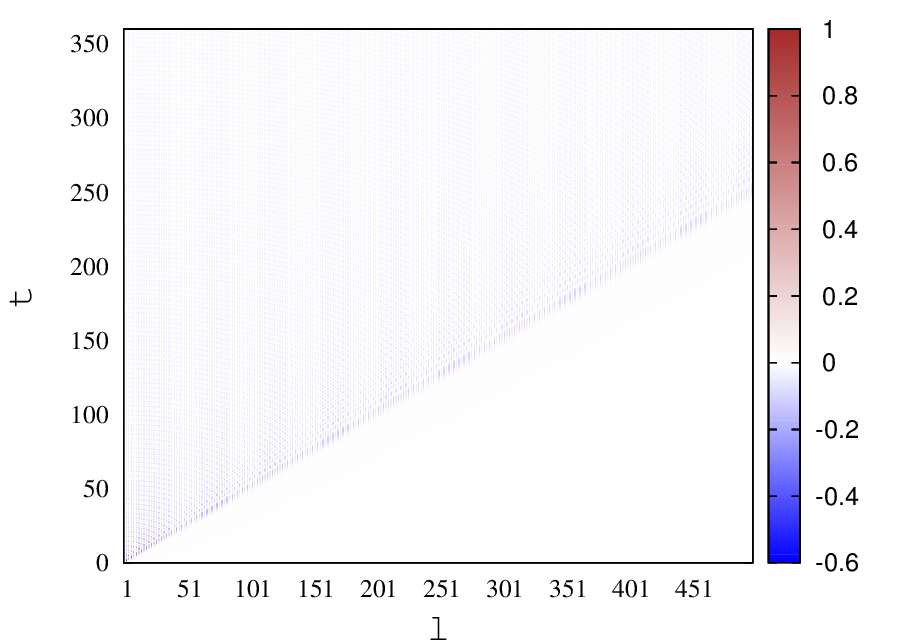}
        }\\
    \end{center}
\caption{\label{fig:fig_2}{Real part of the Green functions $\tilde{G}^x_1(t)$ for the kicked Harper model for Hamiltonian parameters as a function of site index(l) and time (t) (a) $\tau = 0.1$ and $g = 0.1$ (b) $\tau = 0.1$ and $g = 1.0$ (c) $\tau = 0.1$ and $g = 5.0$ (d) $\tau = 0.9$ and $g = 0.1$ (e) $\tau = 0.9$ and $g = 1.0$ (f) $\tau = 0.9$ and $g = 5.0$, (g) for Heisenberg dynamics real part of the Green function $G^x_1(t)$ as a function of site index(l) and time (t).  }
     }%
   \label{fig:subfigures}
   
\end{figure*}

  \begin{figure*}[t]
     \begin{center}
        \subfigure[]{%
            \label{fig:first}
            \includegraphics[width=0.36\textwidth]{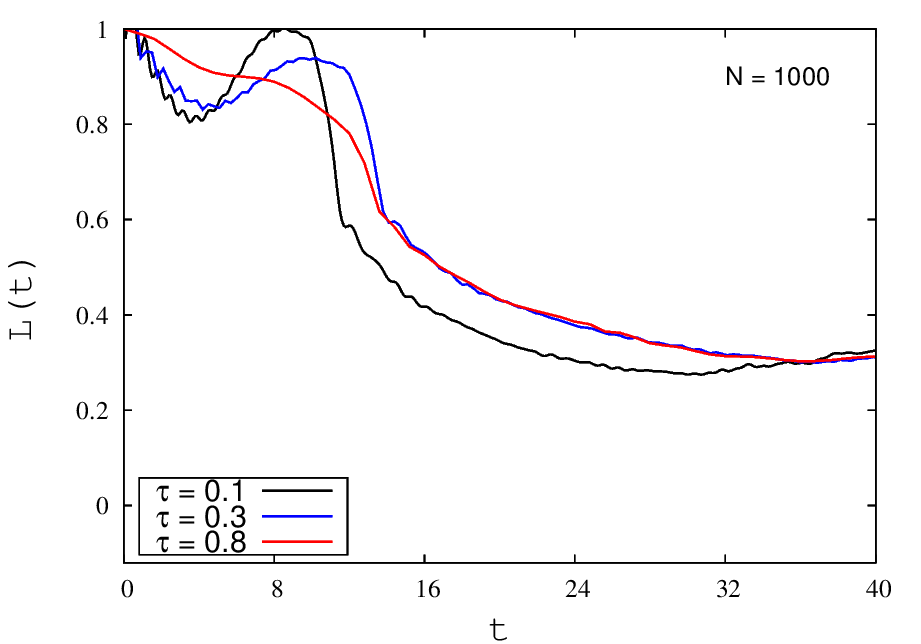}
        }%
        \subfigure[]{%
           \label{fig:second}
           \includegraphics[width=0.36\textwidth]{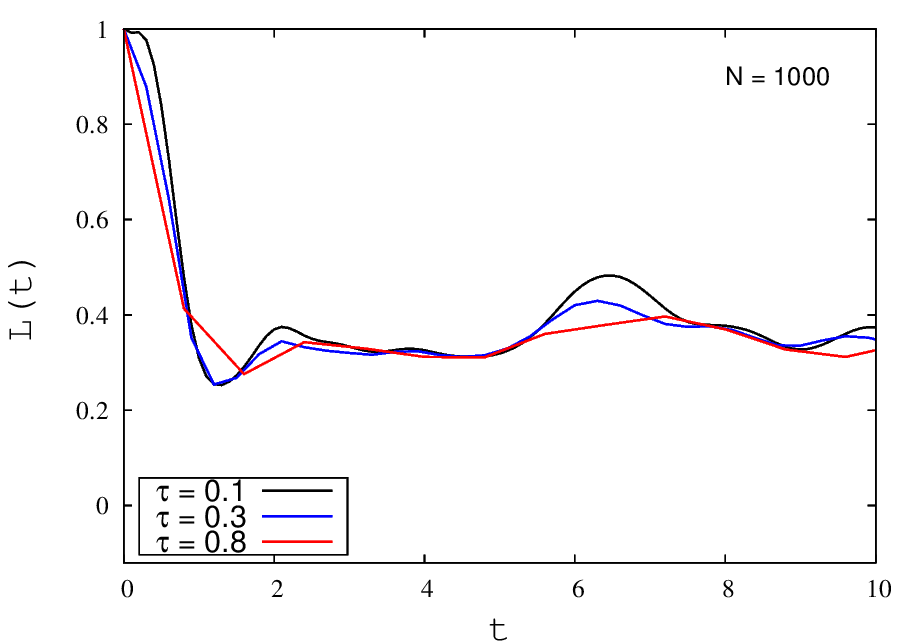}
        }\\
               \subfigure[]{%
           \label{fig:third}
           \includegraphics[width=0.36\textwidth]{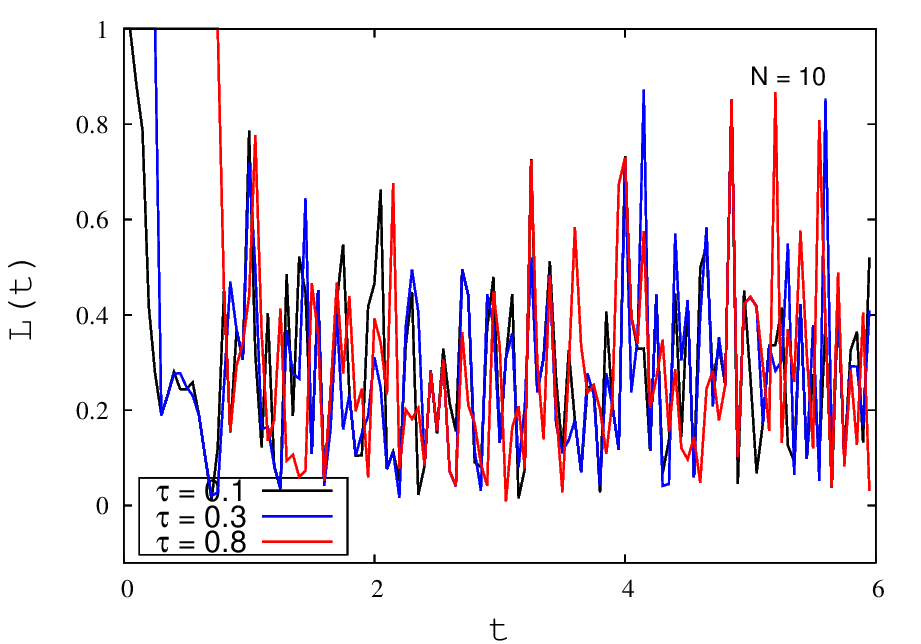}
        }%
        \subfigure[]{%
           \label{fig:fourth}
           \includegraphics[width=0.36\textwidth]{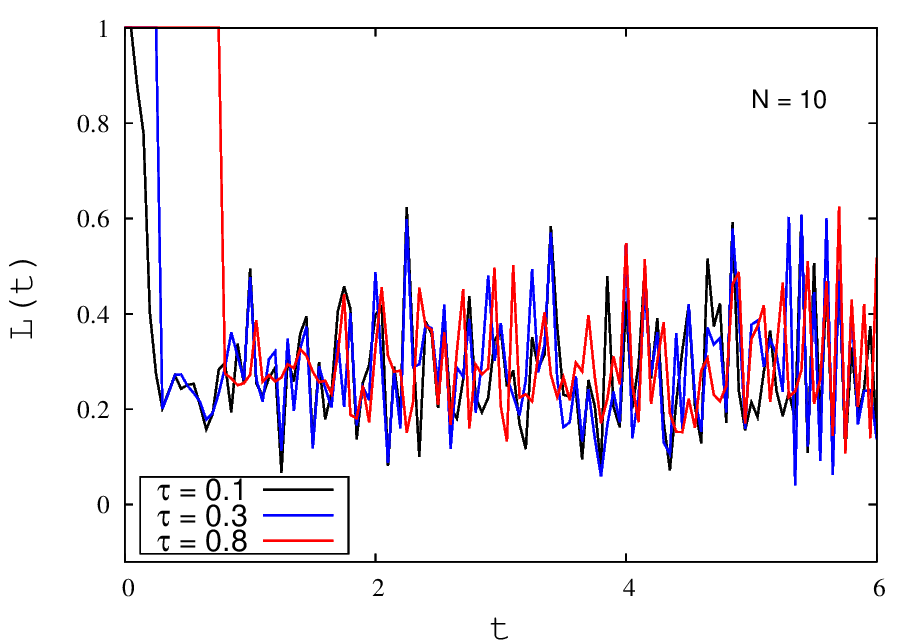}
        }\\
    \end{center}
\caption{\label{fig:fig_2}{ The Loschmidt Echo L(t) as a function of time (t) for Kicked Harper model and XY model for three different values of kicking time $\tau(= 0.1, 0.3, 0.8)$ for potential strength parameter (a) g = 0.1 from analytical calculation,(b) g = 1.0 from analytical calculation, (c) g = 0.1 from numerical calculation,(d) g = 1.0 from numerical calculation. All the numerical calculations has been done for small system size $N = 10$ and analytical results are shown for large system $N =1000$. The initial state taken for numerical calculations is $\frac{1}{\sqrt{2}}| F+ 1\rangle$.  }
     }%
   \label{fig:subfigures}
   
\end{figure*}

  \begin{figure*}[t]
     \begin{center}
       \subfigure[]{%
           \label{fig:third}
           \includegraphics[width=0.36\textwidth]{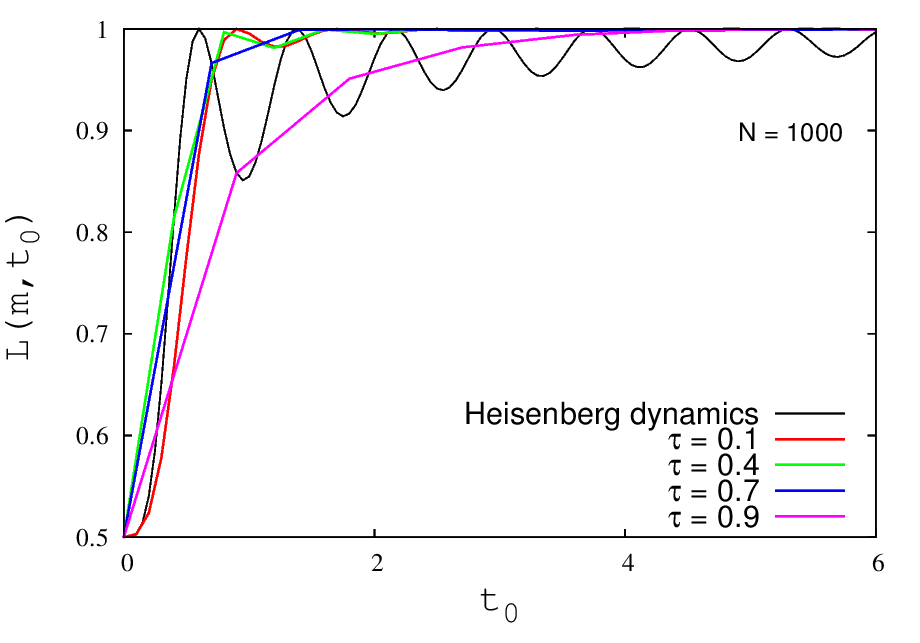}
        }%
        \subfigure[]{%
           \label{fig:fourth}
           \includegraphics[width=0.36\textwidth]{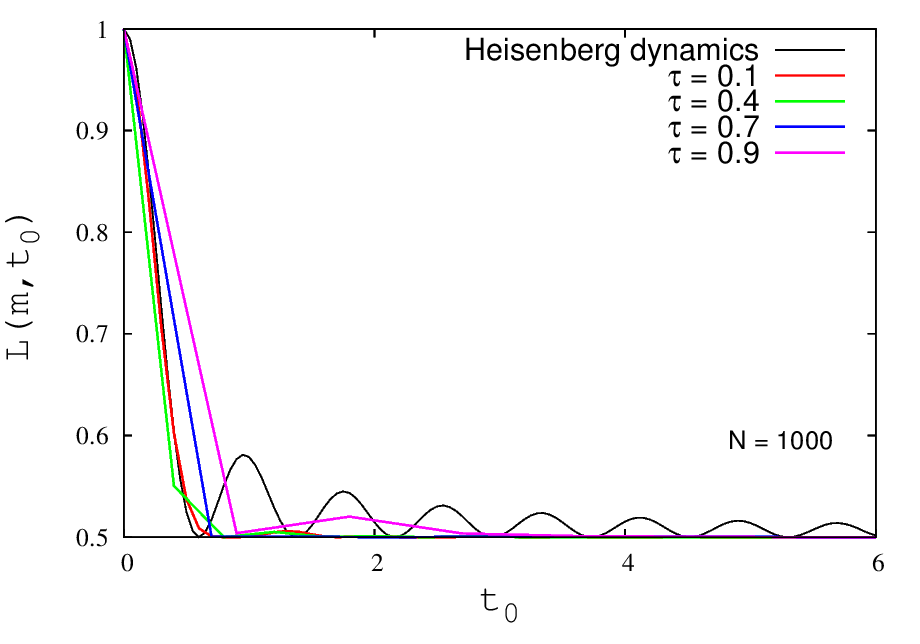}
        }\\%
        \subfigure[]{%
           \label{fig:fourth}
           \includegraphics[width=0.36\textwidth]{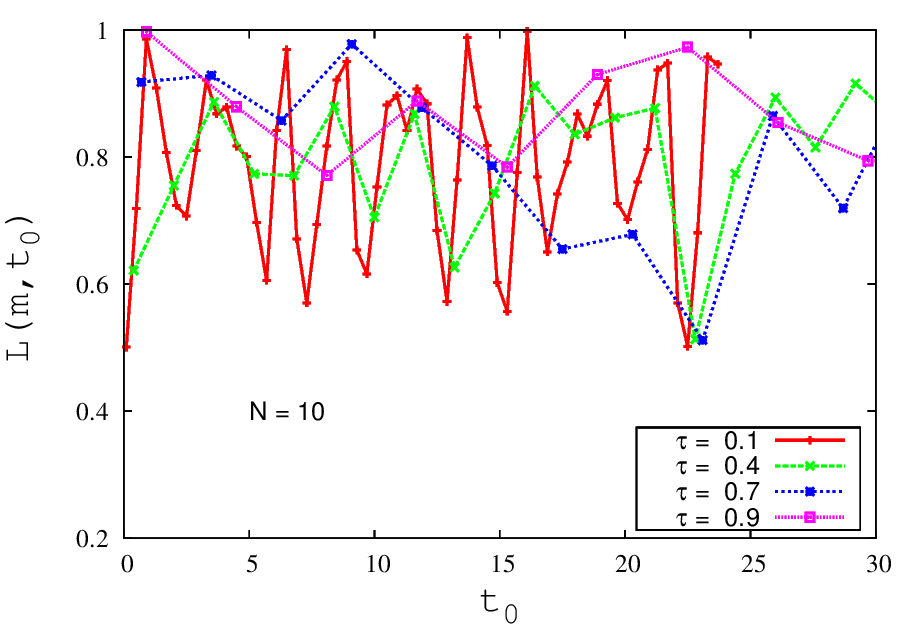}
        }%
        \subfigure[]{%
           \label{fig:fourth}
           \includegraphics[width=0.36\textwidth]{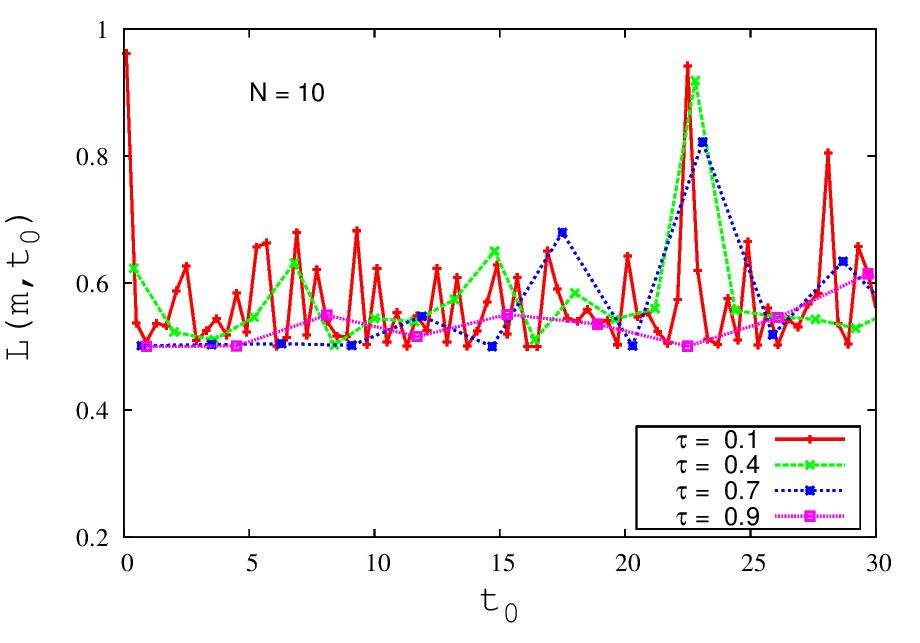}
        }\\%
    \end{center}
\caption{\label{fig:fig_3}{ The Loschmidt Echo $L(t_0)$ for Kicked Harper model as a function of location of the incoherent QDP (projective measurement along $\sigma^z$ basis) ($m$) and the time after which the QDP occurs ($t_0 = n_0\tau$) for four different values of $\tau$ (a) from numerical calculations (b) from analytical calculations. The Loschmidt Echo $L(t_0)$ as a function of location of the incoherent QDP (projective measurement along $\sigma^x$ basis) ($m$) and the time after which the QDP occurs ($t_0 = n_0\tau$) for four different values of $\tau$ (c) from numerical calculations (d) from analytical calculationsIn all the cases the value $g$ is set $1.0$ and $m = 1$.  All the numerical calculations has been done for small system size $N = 10$ and analytical results are shown for large system $N =1000$. The initial state in both the cases is $\frac{1}{\sqrt{2}}| F+ 1\rangle$. }
     }%
   \label{fig:subfigures}
   
\end{figure*}

  By setting $p = 1/2 $ we get the the expression for the Loschmidt Echo for n projective measurements.  Each of the term in the above expression can be computed as the following. Each of the expectation values of multiple operator product of $\sigma^z$ and unitary operator $U$ in the above sum can be written in generally as $\langle \Psi(nt_0)|U_n \sigma^z_{m_n}U_{n-1}\sigma^z_{m_{n-1}}U_{n-2}...U_1\sigma^z_{m_1} U_0|\Psi(0)\rangle$; where, $U_j$ is time evolution operator for time interval $t_j$. For the state    
 $\alpha |F\rangle+ \beta |1\rangle$ the expectation value of such term is given by,
 \begin{eqnarray}
 &\langle \Psi(\sum\limits_{i=0}^{n}t_i)|U_n\sigma_{m_n}^zU_{n-1}\sigma_{m_{n-1}}^zU_{n-2}...U_1\sigma_{m_1}^z U_0|\Psi(0)\rangle = 1+|\beta|^2  [-2\sum^{n}\limits_{j=1}|G^{m_j}_1(\sum\limits_{i=0}^{j-1}t_i)|^2\nonumber\\ &+4 \sum^{n}\limits_{k=2} \sum\limits_{j,k;j<k}^{n}G^{m_j}_1(\sum\limits_{i=0}^{j-1}t_i)G^{m_k}_{m_j}(\sum\limits_{i=j-1}^{k}t_i)G^{*m_k}_1(\sum\limits_{i=0}^{k}t_i)+ ...+(-1)^{n} 2^{n}\prod^n\limits_{j=1}G^{m_j}_1(t_{j-1})G^{*m_n}_1 (\sum\limits_{i=0}^{n}t_i)],\nonumber\\
 \end{eqnarray}
 
 To derive the above expression we have used the following identity for addition of the Green functions,
 \begin{equation}
 \sum_{y'} G^{y'}_{y}(t_1+t_2) G^{*y''}_{y'}(t_1) = G^{y''}_y(t_2).
 \end{equation}

Now, for the entangled initial state  $\alpha |1\rangle + \beta |r\rangle$ the same expression is valid with the quantity $G^{m_j}_1$ replaced by $K^{m_j}$.
Similar to the  above expression, the different terms that arise are: a sum of product of two Green functions, a sum of product of three Green functions, and so on. The contribution becomes smaller as number of Green function increases, as each of them is less than unity. So each of the terms in Eq.16  tends to unity for large value of $t_0$. That means that even if the system is interrupted multiple times at large time intervals $t_0$,  the Loschmidt echo takes a large value. For example, Fig.1(b) shows the value of the quantity $ \langle \Psi(3t_0)|\sigma_{m_3=3}^zU\sigma_{m_{2}=2}^zU\sigma_{m_1=1}^z U|\Psi(0)\rangle$ calculated for the initial state $\alpha |F\rangle +\beta |1\rangle$ taking upto second order, third order and fourth order terms of the Green functions. The contributions becomes much smaller as number of Green functions increases. The Loschmidt Echo $L(\{m_i\};n)$ after multiple number of projective measurement along $\sigma^z$  basis is shown as a function of number of QDPs $n$ in Fig 1(d) for some representative values of $t_0$. The locations of QDPs $m_i$ are some some random number between 1 and 10. Although the Loschmidt Echo is a function of the locations of QDPs, it can be seen that it takes larger values for large  $t_0$ in general. This means that the system gets enough time to revive between two QDPs.  

  \begin{figure*}[t]
     \begin{center}
       \subfigure[]{%
           \label{fig:third}
           \includegraphics[width=0.32\textwidth]{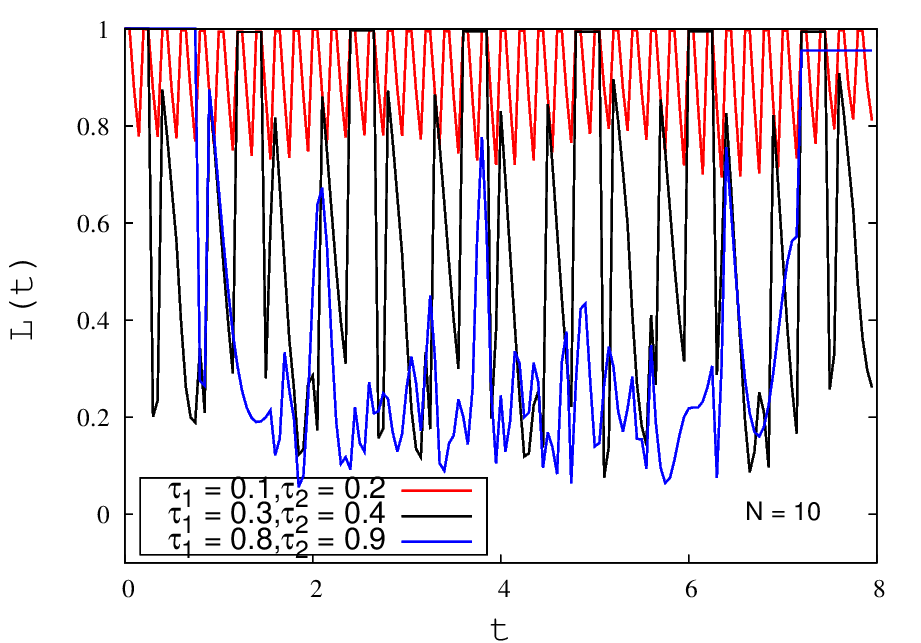}
        }%
        \subfigure[]{%
           \label{fig:fourth}
           \includegraphics[width=0.32\textwidth]{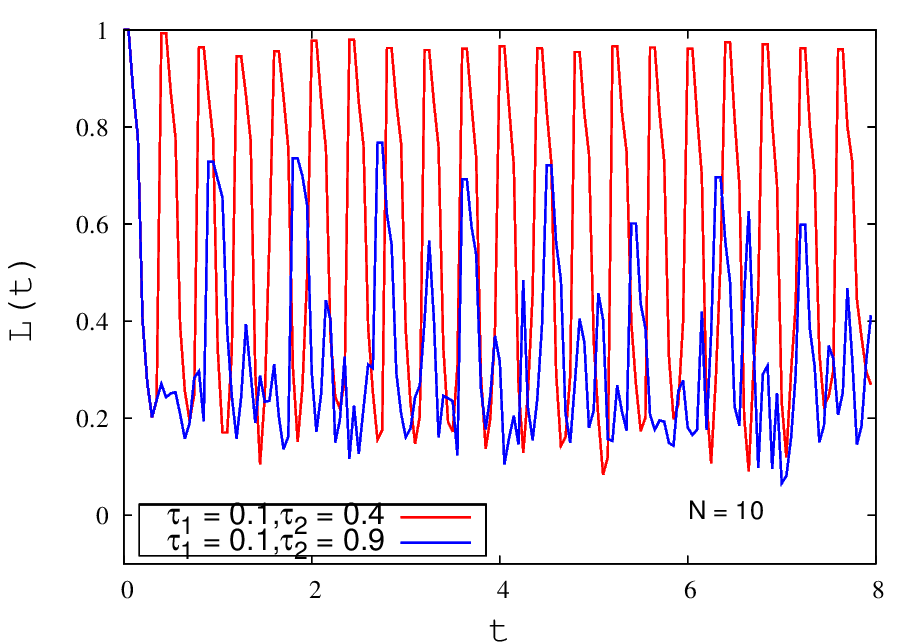}
        }%
        \subfigure[]{%
           \label{fig:third}
           \includegraphics[width=0.32\textwidth]{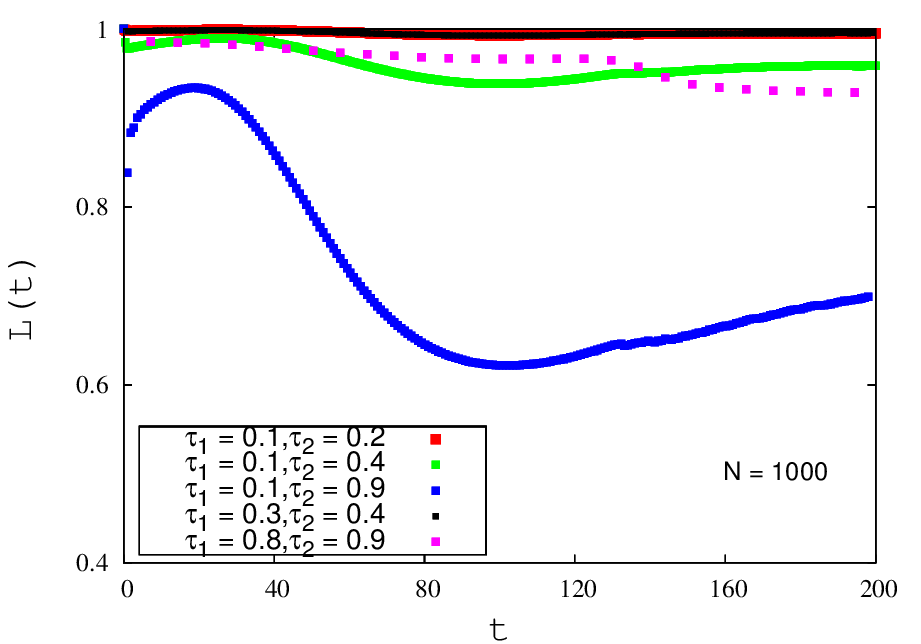}
        }\\%
    \end{center}
\caption{\label{fig:fig_3}{  The Loschmidt Echo $L(t)$ for Kicked Harper model as a function of time (t) for forward kicking period $\tau_1$ and   reverse kicking period $\tau_2$ (a) $\tau_1 = 0.1, 0.3, 0.8$ and $\tau_2 = 0.2, 0.4, 0.9$  from numerical calculations (b) $\tau_1 = 0.1, 0.4$ and $\tau_2 = 0.1, 0.9$ from numerical calculations, (c) $\tau_1 = 0.1, 0.3, 0.8, 0.1, 0.1$ and $\tau_2 = 0.2, 0.4, 0.9, 0.4, 0.9$ from analytical calculations. In all the cases the value $g$ is set $1.0$. All the numerical calculations has been done for small system size $N = 10$ and analytical results are shown for large system $N =1000$. The initial state taken for numerical calculations is $\frac{1}{\sqrt{2}}| F+ 1\rangle$. }
     }%
   \label{fig:subfigures}
   
\end{figure*}

  However, this is not true in case of $\sigma^x$ or $\sigma^y$  expectation values. From similar calculations it can be argued that the $\sigma^x$ or $\sigma^y$  expectation values can be written as sum of individual Green functions each of which tend to zero for large value of $t_0$. Because the operator $\sigma^x$ or $\sigma^y$  changes the number of magnon in the state, in case of conserving dynamics this does not contribute to the expectation value. Let us consider the initial state $\alpha |F\rangle+ \beta |1\rangle$. 
  The operator  $\sigma^x$  acts on this state at  time $t$, generating a state with a linear combination of a zero-magnon state, a one-magnon state and  a two-magnon state.
 The zero-magnon state is generated from one-magnon component of the initial state,  carrying a one-magnon Green's function.  Thus, its overlap with the original state at time $t_0$ tends to zero in the large $t_0$ limit, as shown in Eq 9. This argument can be extended to multiple operations also. Since only zero-magnon  and one-magnon sector 
states contribute to the Loschmidt Echo, these states can come from, at least, one-magnon and two-magnon  states through the $\sigma_x$ operation. So, the expression for the Loschmidt Echo contains the terms like $\sum_x G^{*x}(nt_0)G^{x,x',x'',..}(nt_0)$. This type of terms will eventually go to zero in thermodynamic limit and for $t_0 \rightarrow \infty$, as the $n$-magnon Green function varies with system size as $1/N^{n/2}$. Thus, the value of the Loschmidt Echo saturates to $p^n$, which is the minimum value for large $t_0$ as shown in Eq.15, by replacing the expectation value of $\sigma^z$ by that of $\sigma^x$.

We now discuss the computation of the Loschmidt Echo for a coherent QDP,  which is easier than the incoherent QDP discussed above.  Since the  QDP occurs at $t_0$ during the evolution  the  state $|\tilde \Psi(t)\rangle$ with QDP interruption should match with the state $|\Psi(t)\rangle$ without QDP interruption  except at time $t_0$, $L(m,t)$ should depend on $t_0$ like the previous case. Denoting the instantaneous coherent operation at the site $m$ by a unitary operator $V_m$  the quantity $L(m,t_0)$ in terms of the state at $t_0$ and the instantaneous local gate operator can be written as,
 \begin{eqnarray} 
&{L}(m;t_0)=  |\langle \Psi(t)|\tilde \Psi(t)\rangle |^2 = |\langle \Psi(t_0)|\tilde \Psi(t_0^+)\rangle |^2
=|\langle\Psi(t_0)| V_m|{\Psi}(t_{0})\rangle|^2.
  \end{eqnarray}
In the last step, we have written it as the overlap of the state $|\tilde \Psi(t_0^+)\rangle = V_mU_{t_0,0}|\Psi(0)\rangle$, and the state $| \Psi(t_0)\rangle = U_{t_0,0}|\Psi(0)\rangle$ obtained from the time-reversed unitary process from $t_{0^+}$ to $0$. The dependence of $L(m,t_0)$ on the $m$ and $t_0$ is more transparent, in the above, when it is written as the squared expectation value of $V_m$ in the state $|\Psi(t_0)\rangle$.The operation of the unitary operator $V_m$ on the basis states of $m$'th spin is given by,
\begin{eqnarray}
V_m|0\rangle = \gamma |0\rangle + \delta |1\rangle, V_m|1\rangle = -\delta^*  |0\rangle + \gamma^* |1\rangle. 
\end{eqnarray}
 Here also we take two states viz., one magnon unentangled state $\alpha |F\rangle + \beta |1 \rangle$ and a one magnon entangled state $\alpha |1\rangle + \beta |r\rangle$. The state $\alpha |F\rangle+\beta |1\rangle$ after interrupted by the coherent QDP at time $t_0$ is given by,
 \begin{eqnarray}
  V_m|\Psi(t_0)\rangle =\gamma|\Psi(t_0)\rangle  - 2i \beta\gamma_I G^{m}_1(t_0)|m\rangle +\alpha\delta e^{-i\epsilon_0 t_0}|m\rangle \nonumber\\  + \beta\delta\sum_x G^x_1(t_0)|m,x\rangle-\beta\delta^* G^{m_1}_1(t_0)|F\rangle.
 \end{eqnarray}
 
 The same for the entangled initial state $\alpha |1\rangle + \beta |r\rangle$ is given by,
  \begin{eqnarray}
  V_m|\Psi(t_0)\rangle = \gamma|\Psi(t_0)\rangle  +i \gamma_I( |\Psi(t_0)\rangle -2K^m(t_0)|m\rangle) +\delta \sum_x K^x(t_0)|m,x\rangle - \delta^* K^m(t_0)|F\rangle.\nonumber\\
 \end{eqnarray}
 
 The Loschmidt Echo calculated from the quantity $\langle\Psi(t_0)| V_m|{\Psi}(t_{0})\rangle$ for the state $\alpha |F\rangle + \beta |1 \rangle$ as, 
 
 \begin{eqnarray}
  L(m,t_0) = |\langle \Psi(t_0)| V_m|{\Psi}(t_{0})\rangle|^2 =  |\gamma - 2i\gamma_I|\beta G^{m}_1(t_0)|^2 +2i Im(\alpha\beta^*\delta e^{-i\epsilon_0t_0}G^{m}_1(t_0))|^2. \nonumber\\
  \end{eqnarray}
  The same for the initial state $\alpha |1\rangle + \beta |r\rangle$ is given as,
  \begin{eqnarray}
 L(m,t_0) = |\langle \Psi(t_0)| V_m|{\Psi}(t_{0})\rangle|^2 =  |\gamma - 2i\gamma_I| K^{m}(t_0)|^2|^2.  
 \end{eqnarray}
  
  
  
  Even after the time reversal the state contains zero, one and two magnon sectors. Since the initial state is a combination of zero and one magnon states the two magnon sector does not contribute to the Loschmidt echo. But in case of coherent interrupting QDP  the state obtained after the operation is a combination of  zero, one and two magnon states and each sector has its own conserved dynamics. So, the probability of reviving the system to the initial state is always less than unity.\\
  
  For large value of $t_0$ , the value of $ L(m;t_0) $ saturates to $|\gamma|^2$. For different coherent operations this value  is different depending on the parameters.  For  the X gate  and the Y gate QDPs,  the value is zero ; for the Hadamard gate QDP,  it takes the value $1/2$, and  for the Z gate QDP the value is $1$ which is maximum. The Loschmidt Echo $L(m,t_0)$ as a function of $t_0$ is shown in Fig 1(d) for the two initial states. The value of $\gamma$ and $\delta$ are set as $(1+i)/\sqrt{3}$ and $\frac{1}{\sqrt{3}}$ respectively.  $L(t_0)$ fluctuates around $|\gamma|^2$ which is $2/3$ in this case, for large time $t_0$.
 
 Each time the operator $V_m$ operates on the state the resultant state consists the same state with coefficient $\gamma $. So, we can argue that after n operations the Loschmidt Echo saturates to a value $|\gamma |^{2n}$ for large $t_0$ limit. 
 
 \section{Kicked Harper Model} 
 We now turn our attention to the non integrable dynamics of the the background Hamiltonian evolution.
 There are sharp differences in the eigenvalue spacing distribution and the structure of the eigenfunctions  of the integrable and non-integrable dynamics, that have been widely investigated \cite{lima,arul3}. To see the effect of QDP in non integrable systems we consider a simple model Hamiltonian with a tuneable parameter, to go continuously from completely integrable to completely non-integrable regimes. We use a one-dimensional periodically-kicked Harper model, a simple model of fermions hopping on a chain with an inhomogeneous site potential, appearing as a kick at regular intervals. The spin operator version of the Hamiltonian is given by
 \begin{eqnarray} 
H(t) = \sum_{j=1}^N [-\frac{1}{2}({\sigma}^x_{j}{\sigma}^x_{j+1}+{\sigma}^y_{j}{\sigma}^y_{j+1})+g \sum_{n=- \infty}^{\infty} \delta(\frac{t}{\tau}-n)\cos(\frac{2\pi j \eta}{N}) {\sigma}^z_{j}].
  \end{eqnarray} 
  The first term is the XY term of the Heisenberg model considered above, that causes hopping of up or down spins. The last term is an inhomogeneous magnetic field in the z direction that
  comes into play through kicks at an interval of $\tau$. The coupling strength $g$ and the kicking time $\tau$ can be independently varied, that can affect the nature of the dynamics as
  we will see below.

The classical version of the kicked Harper Hamiltonian is regular for $\tau \rightarrow 0$ and completely chaotic for large value of $\tau$, and similarly the eigenvalue and eigenfunctions of the quantum version display correspondingly a regular or chaotic characteristics\cite{lima}.
The Harper model dynamics conserves the magnon number  through the evolution. So the dynamics can be thought of as site-dependent kicks interrupting the background XY dynamics at a regular interval. Through the time evolution, the down spins can hop around to other sites. We will consider evolution at discrete  times, viz. $t=\tau^+,2\tau^+$ etc, that is at instants just after a kick. The unitary operator for the evolution between two kicks is straightforwardly given by,
 \begin{equation}
 U(g,\tau) = e^{-i \tau \sum_{j} 
-\frac{1}{2}({\sigma}^x_{j}{\sigma}^x_{j+1}+{\sigma}^y_{j}{\sigma}^y_{j+1})}  e^{-i \tau g \sum_j \cos{\frac{2\pi j \eta}{N}} {\sigma}^z_{j} },
\end{equation}
where, the two operator factors appearing above do not commute. The time evolved state at time $n\tau$ just after $n$ kicks is $|\Psi(t)\rangle=U^n(g,\tau)|\Psi(0)\rangle$. The system evolves between a time $n\tau_+$ to $(n+1)\tau_-$ through XY dynamics between two kicks which introduces a lattice position dependent phase factor to the Green function. This is different from the case we have discussed earlier where the background dynamics was interrupted by a coherent QDP at on site; here each kick is a coherent operation which occurs at all sites. We consider $|\Psi(0)\rangle = \alpha |F\rangle + \beta |1\rangle$ as the initial state of the system, the time evolved state will be given by,
 \begin{equation} 
|\tilde{{\Psi}}(t = n\tau_+)\rangle = \alpha e^{-i\epsilon_0 t}|F\rangle + \beta \sum_{x}\tilde{G}^x_1(t = n\tau)|x\rangle.
  \end{equation} 
  
 Here we have introduced a the composite Green function, related to the the Green's function studied in the last section, is given by,
  \begin{equation} 
\tilde{G}^{x_{n}}_{x_0}(t = n\tau) = \sum_{x_1,x_2,...,x_n} \prod^{n-1}_{j=0}{G}^{x_{j+1}}_{x_j}(\tau)e^{2i\tau{ g\cos(\frac{2\pi \eta x_{j+1}}{N})}}.
  \end{equation}  
It can be seen that after each kick, a site-dependent new phase is introduced in the Green function which indicates the qualitative change in the dynamics from the previous section. By setting  $g\tau = 0$ in the above, the Green function $\tilde{G}^{x}_{1}(t)$ it reduces to  the Green function ${G}^{x}_{1}(t)$, the one-magnon propagator function of the
Heisenberg model, as a result of the identity in Eq.17.  The 
real part of the Green function $\tilde{{G}}^{l}_{1}(t = n\tau)$  is plotted for some representative values of $\tau$ and $g$ as a function of site index $l$ and $t$ in Fig 2(a)-2(f). For 
comparison,  the real part of ${G}^{l}_{1}(t)$, that determines the Heisenberg dynamics, is plotted in 2(g). The qualitative nature  depends on the value $g\tau$. $\tilde{{G}}^{l}_{1}(t = n\tau)$ resembles the $G^l_1(t)$  for small value of $g\tau$, this can be seen from 2(a) and  2(d) where the values of 0.01 and 0.09 respectively. Whereas, for larger values  the light cone structure  changes and becomes non linear for larger time, this can be seen from Fig 2(b) and 2(e). For $g\tau$ greater than 0.5 the Green function becomes localised in space and time which can be seen from 2(c) and 2(f). In this range the dynamics is different from the Heisenberg dynamics.

 Since the Loschmidt echo is also sensitive to the dynamics being integrable or non-integrable we investigate the variation of the Loschmidt Echo with time  we take a one magnon initial state and evolve the system unitarily with a Hamiltonian and reverse the system unitarily with a different Hamiltonian. In this context we want to compute the variation of the Loschmidt Echo as a function of number of kicks, kicking interval $\tau$ and potential strength parameter $g$, where in one side evolution there is no kicks  i.e. simple XY dynamics on the other side there is Harper dynamics. The Loschmidt Echo in this case is given by,

  \begin{eqnarray} 
L(t) = |\langle \Psi(t)|\tilde{{\Psi}}( n\tau)\rangle |^2= ||\alpha|^2 + |\beta|^2 \sum_x \tilde{G}^x_1( n\tau)G^{*x}_1(t)|^2.
  \end{eqnarray} 
  
  Fig 3 show the time dependence of the Loschmidt Echo $L(t)$ for different values of $g$ and $\tau$ for both small and large systems. The value of $L(t)$ saturates to the value $1/3$ for large time. This shows that the products of integrable and non-integrable Green functions Eq.29  go to zero for large value of $t$. In case of large systems there is a local revival of the value of $L(t)$ for small values of $g\tau$ and then decays slowly, which can be seen from Fig 3(a). This can be thought of as a signature for non chaotic behaviour. For larger values of $g\tau$ the value of $L(t)$ decays monotonically and fast to the value $1/3$ shown in Fig 3(b). It should be pointed here that the Loschmidt echo fluctuates around 1/3 as shown in Fig. 3(b), after averaging over all possible input states from Eq.28, and it would fluctuate around 1/4 for using a particular set of parameters as shown in Fig. 3(c) and Fig. 3(d). The Loschmidt echo is shown for small systems in Fig. 3(c) and Fig. 3(d), for different values of $g$.  It decays very quickly and then fluctuates around $1/4$, as we are not doing any averaging over all possible input states. The fluctuations are larger for small values of $g\tau$. For smaller systems the value will continue to fluctuate because of repetitive reflections from the boundaries which is explained before.
  
   If an incoherent QDP (e.g. local projective measurement along $\sigma^z$ or $\sigma^x$ basis discussed in section 1) occurs at any site $m$ during the evolution after $n_0$ kicks (i.e. at a time $t_0 = n_0\tau$)  Eq.7 can still be used, with $G^m_1(t)$ replaced by $\tilde{G}^m_1(t)$. Fig.4 shows the time dependence of $L(t)$ for incoherent QDPs. The value of $L(t)$ goes to unity and half for large value of $t_0$ for the case of QDP along $\sigma^z$ basis and $\sigma^z$ respectively, this can be seen from Fig 4(a) and Fig 4(b) respectively. This is similar to the Heisenberg model  except there is no oscillation for larger value of $t_0$ for the Harper dynamics. However, for smaller size systems the behaviour  is similar but $L(t)$ has more fluctuations which is explained earlier. This can be seen in Eq 4(c) and 4(d). In both the cases L(t) shows rapid fluctuations between 0.5 and 1.0. In Fig.4(a), the fluctuations in Loschmidt echo are larger for smaller values of $\tau$, but it often touches unity. But, for larger values of $\tau$, the fluctuations are less, and  the Loschmidt echo is always less than unity. In Fig 4(c) the value of $L(t)$ shows more fluctuations for small values of $\tau$ but mostly bounded between 0.5 and 0.7. We also discussed the case where the system evolves unitarity forward in time with one value of kicking interval $\tau_1$ and in reverse with different value $\tau_2$. The Loschmidt Echo as a function of time in this case in given by,

  \begin{eqnarray} 
L(t =n_1\tau_1 = n_2\tau_2 ) = |\langle \tilde{{\Psi}}(n_1\tau_1)|\tilde{{\Psi}}(n_2\tau_2)\rangle |^2=||\alpha|^2 + |\beta|^2 \sum_x \tilde{G}^x_1( n_1\tau_1)\tilde{G}^{*x}_1(n_2\tau_2)|^2.\nonumber\\
  \end{eqnarray}

The value of the Loschmidt Echo $L(t)$ is shown as a function of time for two different values of $\tau_1$ and $\tau_2$ corresponding to forward and backward evolutions. In smaller size systems, since we evolve the system numerically, $L(t)$ is plotted as continuous function of time $t$ in Fig 5(a) and 5(b). It is seen that the peaks are obtained at time $t = n_1\tau_1 = n_2\tau_2$ where $n_1$ and $n_2$ are integer. For large size system we have calculated the values of $L(t)$ only at the values of $t = n_1\tau_1 = n_2\tau_2$ and plotted in Fig 5(c) where $n_1$ and $n_2$ are integer which corresponds to the peaks in small system. However the qualitative nature is same for both cases. For smaller values of $\tau_1$ and $\tau_2$ e.g. $\tau_2 = 0.1,0.3$ and $\tau_2 = 0.2,0.4$ the peak value is almost unity and remains constant for large $t$. But for larger values of $\tau_1$ and $\tau_2$ e.g $\tau_1 = 0.1$ and $\tau_1 = 0.4$ and $\tau_1 = 0.8$ and $\tau_2 = 0.9$  the peak value is slightly less around 0.95 and slowly vary with time. For  $\tau_1 = 0.1$ and $\tau_2 = 0.9$ the value of $L(t)$ is much less than unity and slowly vary around 0.7 for large time.

\section{Summary and Conclusions}

In this paper, we have investigated the reversibility of a many body quantum state when local quantum dynamical processes interrupt the background dynamics for both integrable and non integrable dynamics. For Heisenberg model we have a one magnon initial state which is interrupted by a local instantaneous operation at a certain epoch of time during evolution. The quantity Loschmidt Echo captures the effect of local operation interrupting the background dynamics. The Loschmidt Echo depends on the time of operation and the Kraus operators corresponding to the operation. It has been shown that the for operations like phase flip operations, projective measurement along $\sigma^z$ the Loschmidt Echo tends to unity for long limit $t_0$. This means that the system loses the memory of the QDP when the QDP occurs after a long time during the evolution or in other words, a long time the quantum state down spin is spread throughout the system, the operation does not change the system considerably. In contrast, for bit flip operation or projective measurement along $\sigma^x$ basis the Loschmidt Echo tends to $p$ or $1/2$ respectively for large $t_0$ limit, as the operation changes the number of magnons in the quantum state. Then we considered multiple sequential QDPs interrupting the background dynamics at different sites at regular interval of time $t_0$.  The expression for Loschmidt Echo in cases can be written a series of sums, where each of the terms can be expanded in terms of product of one particle Green functions, but the leading terms only contribute. There we have shown for the Kraus operators functions of $\sigma^z$ operator the quantum state does not revive to its initial state for finite value of $t_0$ and the value of Loschmidt Echo increases with $t_0$ but decreases with the number of operations $n$. For the operations where the Kraus operators functions of $\sigma^x$ or $\sigma^y$ number of magnon either increases or decreases after each operation. After some number of operations probability of revival to the initial state will become small and for $t_0$ limit tends to zero.\\ \\ We have also discussed the case of local coherent operation on the dynamics. Since such operations also changes the number of magnons in the state, the probability of revival is always less than unity. In case of multiple number of coherent operations a fraction of the initial state is possible to revive. Even for large time interval $t_0$ of operations the value of Loschmidt Echo tends to  value which depends on operation parameter, the value deceases with the number of operations exponentially.\\
\\Finally,  we move to a non-integrable model where an inhomogeneous site dependant potential is added to the integrable XY model. Since the potential is periodic and kicked at regular intervals  this can be thought of as a quantum dynamical process interrupting the system periodically. Each such operation introduces a site dependent phase factor in the Green function. After some number of kicks depending on the potential strength $g$ and kicking interval $\tau$ qualitative nature of the Green function changes. Above certain the value of $g\tau$ the Green function starts to get localised in space and time. This range corresponds to chaos in classical Harper maps.\\ \\In this context we have considered the reversibility of a one magnon state where forward evolution is governed by the kicked Harper Hamiltonian and the reverse evolution is governed by integrable $XY$ Hamiltonian. In such cases the Loschmidt echo shows two different behaviour depending on the value of $g\tau$.  For smaller values, for which the corresponding classical dynamics is non chaotic the Loschmidt Echo decays slowly from unity and shows a temporary revival for a certain time interval for large systems.However, for small size systems though it decays quickly but shows large fluctuations.  Whereas for larger values it always quickly decays with time. We have also considered the case where a local incoherent QDP interrupts the dynamics. In that case the Loschmidt Echo although shows  behaviour similar to integrable models it decays much quickly. As the Green function for the Heisenberg Hamiltonian is basically a Bessel function with time as the argument it oscillates with time which is absent in Harper Green functions as  the value of $g\tau$ increases. Using the same Green function technique we also considered the case where the forward and the backward both the dynamics are governed by same Kicked Harper Hamiltonian but with two different values of kicking periods  $\tau_1$ and $\tau_2$. In this case we see that it decays very slowly and oscillates around a certain value depending on the values of $\tau_1$ and $\tau_2$ instead of going to zero. This value is almost unity for values of $\tau_1$ and $\tau_2$ small. \\
     
 \textbf{Acknowledgement}: SS acknowledges the financial support from CSIR, India.

\end{document}